\documentclass[sigconf]{acmart}

\usepackage{algorithm}
\usepackage{algpseudocode}
\usepackage{amsmath}

\def \Hzero{\mathcal{H}_0}
\def \Hone{\mathcal{H}_1}

\def\PFA{\mathrm{P}_{\mathrm{FA}}}
\def\PD{\mathrm{P}_{\mathrm{D}}}
\def\PSNR{\mathrm{PSNR}}

\def\bx{\mathbf{x}}
\def\bxatk{\mathbf{x}_{\mathrm{atk}}}

\def\bs{\mathbf{s}}
\def\by{\mathbf{y}}

\def\br{\mathbf{r}}
\def\bu{\mathbf{u}}
\def\bm{\mathbf{m}}

\def\key{\mathbf{k}}

\def\beps{\boldsymbol{\epsilon}}

\def\atk{\texttt{atk}}

\AtBeginDocument{%
  }

\setcopyright{cc}
\setcctype{by}
\copyrightyear{2026}
\acmYear{2026}
\acmConference[IH\&MMSec '26]{ACM Workshop on Information Hiding and Multimedia Security}{June 17--19, 2026}{Firenze, Italy}
\acmBooktitle{ACM Workshop on Information Hiding and Multimedia Security (IH\&MMSec '26), June 17--19, 2026, Firenze, Italy}
\acmDOI{10.1145/3785353.3815081}
\acmISBN{979-8-4007-2376-6/2026/06}

\begin{document}
\title{Do Modern Post-Hoc Watermarking Methods Beat Broken-Arrows?}

\begin{abstract}   
With the rapid proliferation of generative models, such as diffusion models, digital watermarking has emerged as a crucial solution for identifying AI-generated images. Modern post-hoc watermarking schemes use neural networks to achieve an extremely low false-alarm rate while remaining robust to common image transformations. 
    However, there is a lack of comparison between these modern methods and classic ones, particularly in real-world scenarios where robustness and security take precedence over achieving an extremely low false-alarm probability.
    In this paper, we propose a fair comparison of robustness and security between modern and classic post-hoc watermarking across various types of classic augmentations and recent sophisticated attacks. 
    Our experiments show that, in a realistic scenario, classic watermarking outperforms modern techniques in terms of security while maintaining robustness. 
\end{abstract}
\author{Enoal Gesny}
\affiliation{%
  \institution{Inria}
  \city{Rennes}
  \country{France}
}

\author{Eva Giboulot}
\affiliation{%
  \institution{Inria}
  \city{Rennes}
  \country{France}
}

\renewcommand{\shortauthors}{Gesny et al.}
\acmArticleType{Research}


\maketitle

\section{Introduction}\label{sec:intro}

With recent advancements in diffusion models~\cite{song2022denoisingdiffusionimplicitmodels, Rombach_2022_CVPR}, generative AI now produces high-quality, diverse, and realistic visuals that are indistinguishable from real images. 
The spread of platforms and services makes this technology accessible to a wide audience. 

This rise in generated images has prompted regulatory entities to react. 
The EU AI Act~\cite{EuropeanAIAct}, the White House executive order~\cite{USAIAnnouncement}, and Chinese AI governance~\cite{ChineseAIGovernance} require that AI-generated content be easily identifiable and traceable.
Among existing methods, such as metadata~\cite{c2pa} and forensics~\cite{corvi2023detection}, digital watermarking stands out as a key approach to address the issue. 

This leads to a revival of interest in the design of new watermarking techniques. Classical methods modified Fourier or wavelet coefficients~\cite{cox_digital_2008, furon:hal-00335311} to leverage the theoretical literature on watermarking designs~\cite{costa_writing_1983, miller_informed_2000}. This approach combined knowledge from signal processing~\cite{pateux_practical_2003}, information theory~\cite{merhav_optimal_2006}, and statistical theory~\cite{miller_computing_2000} in order to inform their design. The modern approach, on the other hand, is now fundamentally based on machine learning, leveraging the flexibility of deep neural networks~\cite{ferrari_hidden_2018, bui_trustmark_2023, fernandez_video_2024}. 

The need to trace AI-generated content spurred a line of research specifically dedicated to take advantage of the generation process itself~\cite{yang2024gaussian,fernandez2023stable,gesny2025guidancewatermarkingdiffusionmodels}. However, the difficulty of integrating these methods into existing infrastructure has led the industry to favor post-hoc methods~\cite{bui_trustmark_2023,fernandez_video_2024,gowal2025synthidimageimagewatermarkinginternet}. 
Post-hoc methods embed the watermark in content after it has been generated, meaning they cannot reliably be used for open-source/open-weights workflows. This is of no consequence for proprietary models: the generation pipeline is kept behind an API. In this setting, modern post-hoc methods offer a plug-and-play approach with detection performance that can easily be tuned.

What we term the \textit{modern approach} in this paper can be characterized by three main ingredients: 1) a multi-bit watermarking framework, 2) an embedder and a decoder designed as deep-neural networks, and 3) a training pipeline that explicitly includes robustness and quality considerations by introducing image-processing augmentations as well as psycho-visual masking~\cite{chou_perceptually_1995}. The simplicity and flexibility of this recipe allowed the rapid refinement of embedders and decoders since the seminal HiDDeN architecture~\cite{ferrari_hidden_2018}.

Nevertheless, the modern approach vastly increases the attack surface on the watermarking channel. First of all, they inherit the vulnerability of deep-neural networks to adversarial examples~\cite{goodfellow_explaining_2015}, making detectors extremely brittle against white and black-box attacks~\cite{imadache_evaluating_2025}. Secondly, modern detectors lack the concept of the \textit{watermarking key}. Combined with the first vulnerability, this means that the watermarking channel is fully compromised as soon as the detector is made public, since erasure and copy attacks are then trivial to perform. This is fundamentally a \textbf{security} problem, which is vastly more understood and studied in the classical approach~\cite{cayre_watermarking_2005-1,furon_new_2012,bas_watermarking_2016}. 

Another arresting decision with the modern approach is its focus on multi-bit watermarking. In practice, content traceability does not necessitate decoding a message. Indeed, re-framing watermarking as a detection problem -- i.e. zero-bit watermarking -- is a proven way to tackle the traceability problem~\cite{rhoads_detecting_2010, chappelier_procede_2018}. We further discuss the question of multi-bit and zero-bit watermarking in Section~\ref{sec:zerobit}. For now, it suffices to observe that modern multi-bit decoders can be converted to zero-bit detectors easily -- see~\cite{gesny2025guidancewatermarkingdiffusionmodels}, which demonstrates the excellent performance of such conversion.  

Stemming from these observations, this work aims to challenge the assumed superiority of the modern approach by asking the following question: \textbf{Does modern post-hoc watermarking consistently outperform classical approaches in a zero-bit setting?} 
We answer this question through the following contributions: We establish a framework to compare methods and attacks under zero-bit assumption, we provide a fair comparison of modern and traditional watermarking, and we compare recent attacks within the same realistic framework.

\section{Zero-Bit Watermarking}\label{sec:zerobit}

Zero-bit watermarking is fundamentally different from multi-bit watermarking: the former is a detection problem, whereas the latter is a communication problem. We frame it as a hypothesis test between two hypotheses: $\Hzero$ the image is not watermarked, $\Hone$ the image is watermarked. Zero-bit watermarking aims to embed a signal in the host such that the power $\PD$ of the detector is maximized under a guaranteed, fixed, probability of false-alarm $\PFA$.

The watermarked detector is a function $\phi : \mathbb{R}^{D} \times \mathcal{K} \rightarrow \mathbb{R}$ that takes an image $\bx \in \mathbb{R}^{D}$ and a key $k \in \mathcal{K}$ as input. The test $\delta_k(\bx)$ then decides on either hypothesis by comparing the detector's output against a fixed threshold $\tau \in \mathbb{R}$:
\begin{equation}\label{eq:stat-test}
	\delta_k(\bx) := \phi(\bx, k) \lessgtr^{\Hzero}_{\Hone} \tau.
\end{equation}

In practice, the distribution of the detector's output is often known under $\Hzero$. It is then simpler to work directly with the $p$-values $p_{\phi}(\bx,k) := F^{-1}_{\phi}\left(\phi(\bx,k)\right),$
where $F^{-1}_{\phi}$ is the quantile function associated with the distribution of $\phi(\bx,k)$. 

When $F_{\phi}$ is absolutely continuous, the distribution of the $p$-values is uniform. Replacing $\phi$ by $p_{\phi}$ in Eq.\eqref{eq:stat-test}, one guarantees a level $\alpha \in [0,1]$ for the test simply by setting $\tau = \alpha$.


\subsection{The hypercone detector}

Let $\key$ be a normalized secret vector in a (secret or not) $M$-dimensional subspace $\mathcal{S}_{\key} \subset \mathbb{R}^{M}$. We can build a detector $\phi$ in two steps. Let $\bx \in \mathbb{R}^{D}$ be an input image and $P_{\key}: \mathbb{R}^{D} \rightarrow \mathcal{S}_{\key}$ the projection to the key subspace. We can compare the extracted vector $\br := P_{\key}(\bx)$ to the secret vector $\key$ by computing the angle $\theta$ between the two, $c(\br, \key) := \frac{|\br^{T} \key|}{||\br|| } = \cos \theta$.


Now, if we construct $P_{\key}$ such that the distribution of $P_{\key}(\bx)$ is isotropic under $\Hzero$, the probability that $\br$ falls inside the hypercone of axis $\key$ and half-angle $\theta$ is given by:

\begin{equation}\label{eq:hypercone-pfa}
	\PFA = 1-I_{\cos^2\theta}\left(1/2, (M-1)/2\right).
\end{equation}

This is indeed the probability of false alarm of the test based on the detector $\phi_{\mathrm{hypercone}}$ defined as $\phi_{\mathrm{hypercone}}(\bx,\key) := c\left(P_{\key}(\bx), \key\right).$

\subsection{Applications}
\paragraph{Broken-Arrows} In the original work~\cite{furon:hal-00335311}, the authors use $\phi_{\mathrm{hypercone}}$ as the detector. Both the secret vector $\key$ as well as the projection $P_{\key}$ are secret. The projection is composed of two steps. The first step performs a (public) wavelet transform on the input and conserves the first $N_f$ coefficients. These coefficients are then projected into the secret subspace using $M$ secret (pseudo)-orthogonal basis vectors. Within this subspace, $N_c$ hypercones are defined, using the closest one to the host is selected for embedding: this is the secret vector $\key$. The host vector is then pushed as far as possible from the detection border in order to maximize robustness. Note that due to the use of several hypercones, the $\PFA$ is slightly higher than reported in Eq.~\eqref{eq:hypercone-pfa}. Using a simple union bound, the $\PFA$ for Broken-Arrows is such that $\PFA^{(\mathrm{BA})} \leq N_c \PFA$.

\paragraph{Modern multi-bit approach} The hypercone detector can be trivially applied to modern multi-bit watermarking by observing that their decoder are all based on projecting an input into a $M$ dimensional subspace, $M$ being the number of bits in the message. The message is then decoded by thresholding the value of each vector element, usually assigning $0$ and $1$ to negative and positive elements respectively. By skipping the thresholding step, we obtain the projection $P_{\key}$ for free. It remains to define the secret vector $\key$. To do so, notice that the original multi-bit schemes embed a message $\bm \in \{0,1\}^{M}$. We can construct the secret vector by modulating $\bm$ with antipodal modulation: we map 0s to -1s and 1s to 1s. Finally, we normalize the modulated message to obtain the final secret vector $\key \in \{-1/\sqrt{M}, 1/\sqrt{M}\}^M$. The isotropy assumption must be checked carefully for this approach -- see \cite{gesny2025guidancewatermarkingdiffusionmodels}[Appendix C] for a generic methodology to enforce this assumption.

\section{Evaluation methodology}

The recent watermarking reference~\cite{bas_watermarking_2016} defines a watermarking system as:
\textit{``[...] the embedding of a robust, imperceptible and secure information.''}. 
These properties were given thorough definitions in \cite{kalker_considerations_2001}:
\textit{``Robust watermarking is a mechanism to create a communication channel that is multiplexed into an original content [where] the perceptual degradation of the marked content [...] with respect to the original content is minimal and [where] the capacity of the watermark channel degrades as a smooth function of the degradation of the marked content.''} and \textit{``Watermark security refers to the inability by unauthorized users to have access to the raw watermarking channel.''}.

We herein propose a methodology to compare and rank the performance of each watermarking based on these three axes. We motivate our decision in each case, with the goal to design an evaluation which is \textit{fair} and \textit{relevant to a realistic implementation of watermarking systems}.


A first important choice is the nature of images to use. Experiments are conducted on $1024 \times 1024$ natural images to align with the scale-dependent capacity of watermarking and the high-resolution standards of generative models and digital media.

\subsection{Detectability/Robustness}

We can rank the robustness of watermarking schemes according to two main methods. The first is by computing some statistics on the $p$-values of each scheme across different attack scenarios. The second is by setting a detection level $\alpha$ guaranteeing a chosen $\PFA$. In practical scenarios, a detection threshold is always fixed, we thus focus on the second method. The hypothesis test is thus simply: $p(\bx) \lessgtr^{\Hzero}_{\Hone} \alpha$.

How low should we set the level $\alpha$? We can observe a race towards higher capacities in modern watermarking schemes, with TrustMark~\cite{bui_trustmark_2023} starting at 100 bits, followed by Videoseal~\cite{fernandez_video_2024} at 256 bits, and most recently ChunkySeal~\cite{petrov2026we} at 1024 bits. When no attack is performed, more capacity usually translates to higher detectability after converting into a zero-bit detector. In a real-world scenario, we argue that guaranteeing extremely low $\PFA$, say $10^{-100}$, under a few image processing attacks is far less valuable than guaranteeing a modest $\PFA$, say $10^{-6}$, across all possible attack scenarios.

\paragraph{Decision} We follow the BOWS-2 competition~\cite[Section 6]{furon:hal-00335311} and set $\alpha = 10^{-6}$ so that in expectation, one image out of one million is a false positive. 
\subsection{Imperceptibility}

In this work, we compare the quality between two images using the standard choice of the $\PSNR(\bx, \by) = -10 \cdot \log_{10} \left( \frac{1}{D}\sum_{i=1}^{D}(x_i - y_i)^2 \right).$
 We are aware of the use of perceptual metrics such as the LPIPS~\cite{zhang_unreasonable_2018} and psycho-visual masks (JND)~\cite{chou_perceptually_1995} during the training of modern watermarking encoders. However, in order to be fair with respect to detectability, we prefer working at a fixed watermark signal power which makes the $\PSNR$ a natural choice in this case. 
 
 To ensure fairness, we thus fix the $\PSNR$ of watermarking signal to a fixed level $\eta$ by scaling the signal after embedding such that:
 
 \begin{equation}
 	\bs := \psi(\bx) - \bx ;\quad
 	\bx_{\mathrm{wm}} = \bx + \bs/||\bs||\sqrt{D}10^{\frac{\eta}{20}}.
 \end{equation}
 
 
 \paragraph{Decision} We measure the impact of watermarking on image quality using the PSNR between the original and the watermarked images. We fix $\eta=$42dB. We measure the impact of an attack using the PSNR between the watermarked and the attacked images.



\subsection{Security}\label{subsec:method-security}

We now propose four scenarios where the attacker's goal is to erase the watermark of the image. More precisely, an attack is deemed successful if the attacked image $p(\bxatk) < \alpha$, that is, if its p-value is below the level $\alpha$ of the detector.
We present each scenario in decreasing order of knowledge of the attacker about the detection process.

\paragraph{White-box scenario} We provide the attacker with all the knowledge pertaining to the detection process. For the modern approach, this corresponds to access to the detector as a white-box. This most closely reflects the standard Kerchoff's principle, where everything should be public but the key. 

For the modern approach, we formulate the problem as the optimization of an adversarial perturbation:
\begin{equation}\label{eq:adv-opt}
	\begin{cases}
		\min_{\beps\in\mathbb{R}^D} &||\beps||^2_2\\
		\mathrm{s.t.} &p(\bx + \beps) > \alpha
	\end{cases}
\end{equation}
This is usually solved iteratively, with a fixed model query budget of $Q$.
For this paper, we settled on the boundary projection \textit{DDN Attack}~\cite{ddnattack} with gradient.
 
For Broken-Arrows, we also disclose the key to the attacker. Though unfair to Broken-Arrows, this allows us to use closed-form formulas to compute the optimal attack using Eq.(11) and Eq.(12) in~\cite{furon:hal-00335311}.

\paragraph{Black-box scenario} The attacker only has access to the detector as a black box: they can query it and obtain a yes/no answer, but the gradient cannot be computed exactly through backpropagation. This attacker still aims to solve the optimization problem in Eq.~\ref{eq:adv-opt}. We settled on CGBA~\cite{Reza_2023_ICCV}, which is currently the most efficient black-box attack for locally linear classifiers.

\paragraph{Oracle scenario} Contrary to previous scenarios, the attacker does not have access to the detector. They instead use an oracle $O : \mathbb{R}^D \rightarrow \mathbb{R}$ which serves as a proxy to the detector's output. We selected two attacks as representative of this scenario, with two different approaches. The first, WmForger~\cite{soucek2025transferableblackboxoneshotforging}, uses a preference model as an oracle and performs a gradient ascent on the image. In other words at the $i$-th iteration, we have:
\begin{equation}\label{eq:wmforger}\tag{WmForger}
	\bxatk^{(i)} = \bxatk^{(i-1)} - \nabla_{\bxatk^{(i-1)}}O\left(\bxatk^{(i-1)}\right).
\end{equation}
The attack stops after a fixed number of iterations $Q$.
The second attack, Watermark In the Sand (WIS)~\cite{wm_in_the_sand}, uses the oracle as a stopping condition. It locally modifies the image at each step and continues until the condition is reached. We found the design in the original paper -- based on inpainting and GPT-as-a-judge oracle for quality -- computationally intensive and inefficient. We simplified the design for this paper, purifying local patches using a VAE and using the PSNR as an oracle. This resulted in far better attack performance at a fraction of the computational cost -- see Appendix~\ref{app:wits} for details about our implementation.

\paragraph{Blind scenario} The attacker has no access to the detector and uses no feedback from an oracle. This is the classic robustness scenario in watermarking that tests detectability against sets of classic image processing transformations -- e.g. JPEG compression, sharpening, gamma transform, etc. However, recent attacks based on the use of diffusion models and VAEs, such as Purification~\cite{nie2022DiffPure}, also fall in this category since the attacker must set the number of steps blindly. 
For this paper, we made the decision \textbf{to focus exclusively on valuemetric operations}. The reason is that the robustness against geometric operations such as crops, rotations, perspective shifts, etc, can be generically improved by decomposing the image into patches -- this is the approach recently taken by SynthID~\cite{gowal2025synthidimageimagewatermarkinginternet} -- and synchronization techniques~\cite{fernandez2025geometricimagesynchronizationdeep}\cite{cox_digital_2008}[Section 9.3]. Since we want to focus on the performance of the baseline watermarking systems at a fixed image size, we forego the comparisons to these operations.

\section{Results}
\subsection{Experimental Settings}

\paragraph{Dataset} We perform watermark embedding and attacks on 200 natural images from the  MFlickr~\cite{flickr} dataset. All images were resized to $1024 \times 1024$ using bilinear interpolation, with further cropping applied to preserve the aspect ratio.

\paragraph{Modern watermarking} We chose Videoseal~\cite{fernandez_video_2024} and TrustMark (without error correcting codes)~\cite{bui_trustmark_2023} as modern representatives of post-hoc methods. They were converted to a zero-bit watermarking scheme using the methodology in Section~\ref{sec:zerobit}. The detectors were whitened using the methodology presented in Appendix C of ~\cite{gesny2025guidancewatermarkingdiffusionmodels} in order to match the isotropy assumption of the hypercone detector. The dimensions of the watermark subspace for Videoseal and TrustMark are $M=256$ and $M=100$ respectively.

\paragraph{Classical watermarking} We chose Broken-Arrows~\cite{furon:hal-00335311} as the state-of-the-art of the classical approach. We perform a three levels wavelet decomposition on the GPU using the \textit{Pytorch Wavelet Toolbox} library\footnote{see \url{https://github.com/v0lta/PyTorch-Wavelet-Toolbox}} with Daubechies-9 wavelets. We embed only within the first $N_f=60492$ coefficients. We set the dimension of the subspace as $M=128$ and use $N_c=50$ secret hypercones. 

For all methods, the key is randomized for each image, and the watermarking signal is scaled to achieve a fixed PSNR of 42dB with respect to the original image. 

\paragraph{White-box scenario} We performed the DDN Attack~\cite{ddnattack} on modern watermarking techniques with a fixed query budget $Q=250$. 
For Broken-Arrows, we perform the optimal attack as described in Section~\ref{subsec:method-security}.
\paragraph{Black-box scenario} We perform the state-of-the-art CGBA attack~\cite{Reza_2023_ICCV} for both modern and classic watermarking with a fixed budget of $Q=2000$ queries. We restrict the search space by keeping only the lowest $1.25\%$ DCT frequencies (including DC) -- see Section 5.2 in \cite{maho_surfree_2021}. Preliminary experiments showed an average search overhead of $10$ queries to find the adversarial point at each iteration, irrespective of the method. Following the theoretical recommendations in~\cite{gesny_when_2024}, we thus set the number of queries for estimating the gradient to $2$.
\paragraph{Oracle scenario} We use two novel methods: WmForger~\cite{soucek2025transferableblackboxoneshotforging} and our version of the Watermark In the Sand (WIS) attack as defined in Appendix~\ref{app:wits}. For the latter, the image is downsampled to $512\times 512$ before being passed through the VAE. We use the SANA~\cite{xie2024sana}\footnote{Huggingface ID: \url{Efficient-Large-Model/Sana\_600M\_512px_diffusers}}.

\paragraph{Blind scenario} We use the SANA and Stable-Diffusion-2.1\footnote{Huggingface ID: \url{stabilityai/stable-diffusion-2-1-base}} diffusion models for purification. The Flow-Matching Euler Discrete scheduler and Euler Discrete schedulers are used respectively. For basic image processing valuemetric operations, we only succeeded to attack the detectors with JPEG compression at a quality factor of 5. 

\begin{figure*}
    \centering
    \includegraphics[width=1\linewidth]{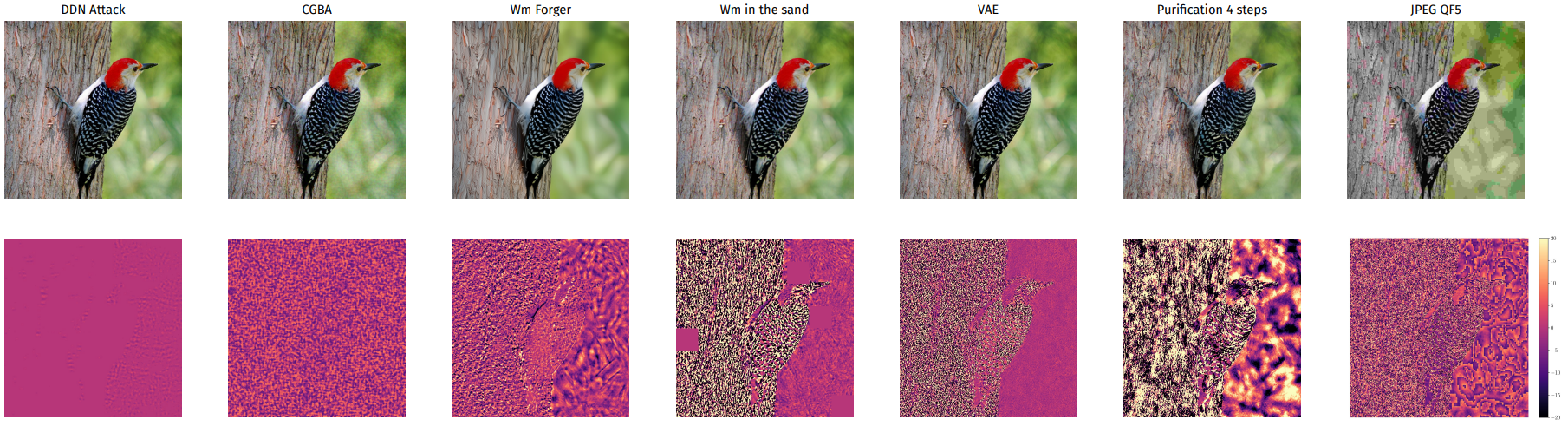}
    \caption{Examples of attacks on Videoseal.}
    \label{fig:examples}
\end{figure*}

We reported examples of all used attacks with their residue compare to the watermarked image in figure~\ref{fig:examples}.

\paragraph{Metrics} We compute the attack success rate of each attack as $ASR = \frac{\sum_{i=1}^N 1(p_\phi(\atk(\bx_i), k_i) > \alpha)}{N}$,
where $N$ is the number of images $x_i$ and $\atk$ is the evaluated attack.  
The threshold for attack success is set at $10^{-6}$. We measure the distortion due to the attack  using the PSNR \textbf{with respect to the watermarked image}.


\subsection{Experimental Results}

\begin{figure}[t]
    \includegraphics[width=0.45\linewidth]{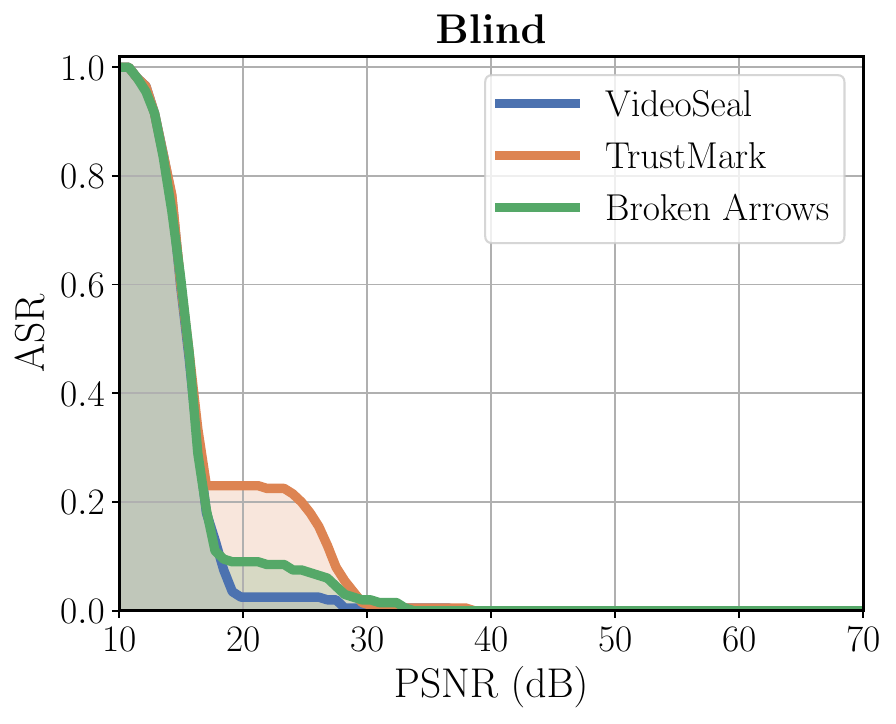}
    \includegraphics[width=0.45\linewidth]{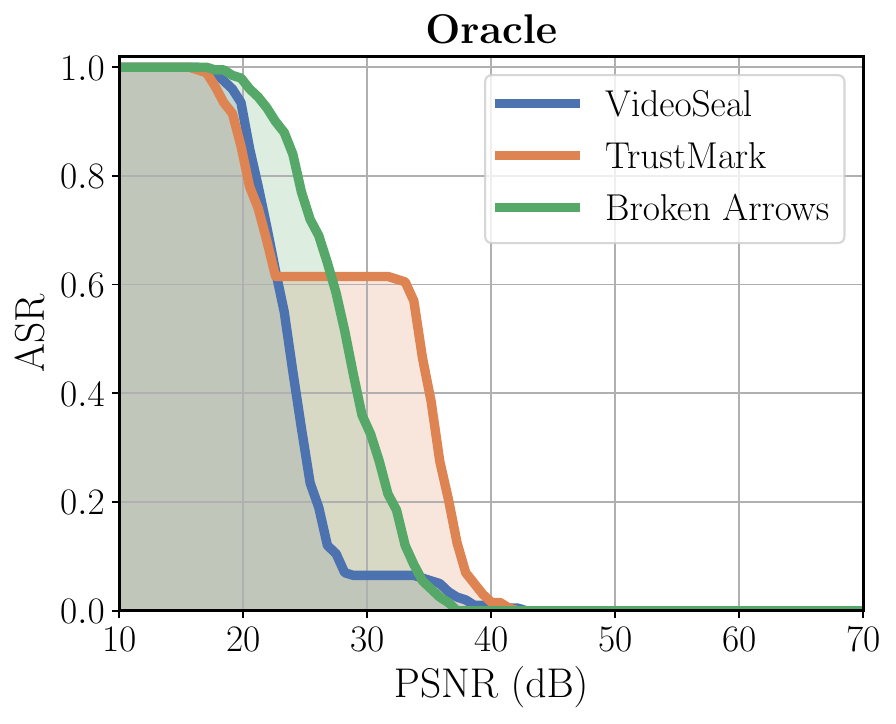}
    \includegraphics[width=0.45\linewidth]{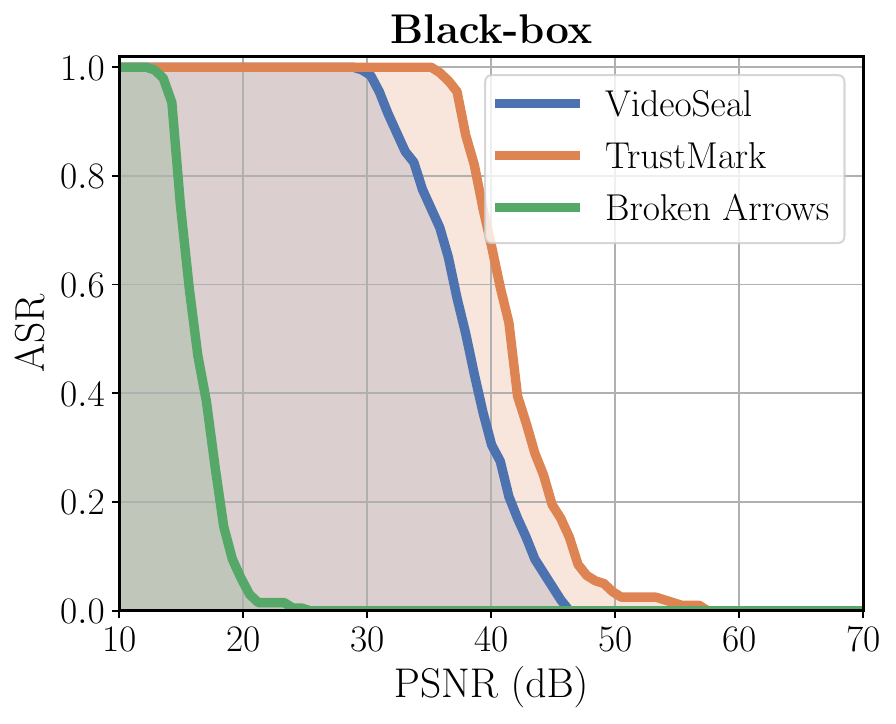}
    \includegraphics[width=0.45\linewidth]{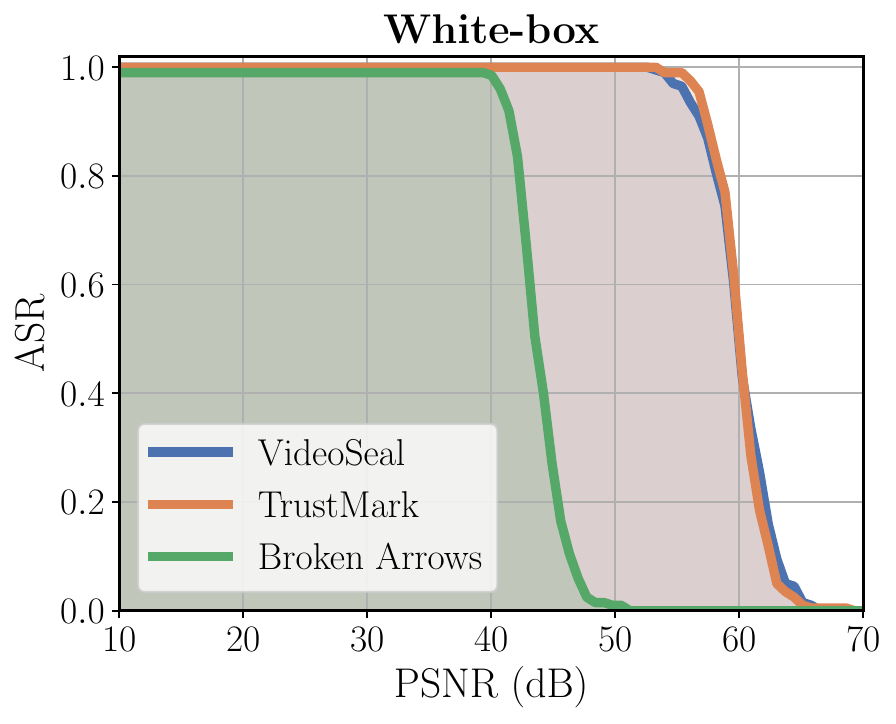}
    \caption{Comparison of watermarking schemes across scenarios. Each curve is the worst-case attack envelope computed from ASRs of attacks corresponding to the same scenario as a function of the PSNR. A smaller area is better for the watermark.}
    \label{fig:scenario}
\end{figure}

We report the ASR at a threshold of $\alpha=10^{-6}$ as a function of PSNR for each scenario and watermarking technique in figure~\ref{fig:scenario}.
\paragraph{White-box scenario} Broken-Arrows clearly outperforms TrustMark and Videoseal. For the latter, the attack achieves a distortion nearing the sub-quantization limit around 60dB, whereas the true optimal attack for Broken-Arrows achieves $100\%$ only around 40dB. 
Due to their finite training set, the detection region of DNN-based detectors in pixel space contains many "blind spot" making them vulnerable to small perturbations~\cite{goodfellow_explaining_2015}. 
The possibility of computing gradients makes these blind spots easy to identify with a gradient projection attack. The way Broken-Arrows detection is built precludes the existence of such adversarial signals

\begin{figure}
    \centering
    \includegraphics[width=1\linewidth]{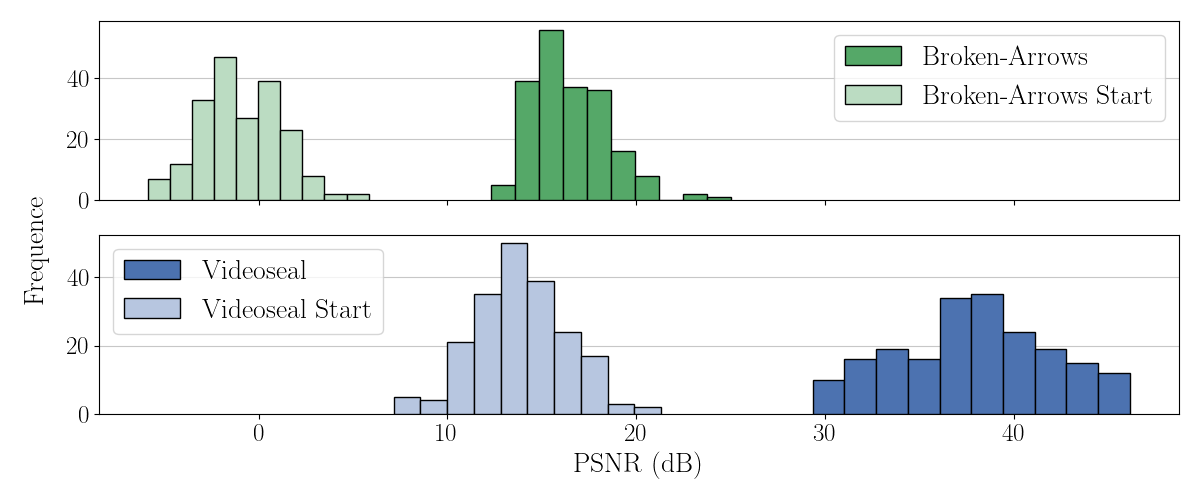}
    \caption{Distribution of the PSNR of the images attacked after 10 queries and 2000 queries of CGBA against Videoseal and Broken-Arrows.}
    \label{fig:distrib_psnr}
\end{figure}

\paragraph{Black-box scenario} Both Videoseal and TrustMark are once again far more vulnerable than Broken-Arrows. The gap between the modern and classical methods is surprisingly large: nearly 15dB separates the $100\%$ ASR threshold between the two approaches. The reason for the lack of success against Broken-Arrows is surprising as it fully meets the assumptions of the attack, notably the linearity of the detector. From the results in the white-box scenario, we have an upper-bound on the smallest distortion for a successful attack. Leveraging the theoretical study in~\cite{gesny_when_2024}, we can estimate the expected distortion at each attack iteration depending on the distortion after one iteration. The main insight from this analysis is that if finding a good adversarial example is difficult early on, convergence to the optimum will be slow.
Figure~\ref{fig:distrib_psnr} presents the PSNR distribution for the initial adversarial points and their state after 2000 queries across both watermarking techniques.
The results indicate that initial boundary searches against Broken-Arrows incur a higher distortion cost compared to Videoseal.
We hypothesize that Videoseal susceptibility to find a better initial boundary point in the pixel space arises from the inherent geometry of DNN detectors. 

\begin{figure}[h]
    \includegraphics[width=0.45\linewidth]{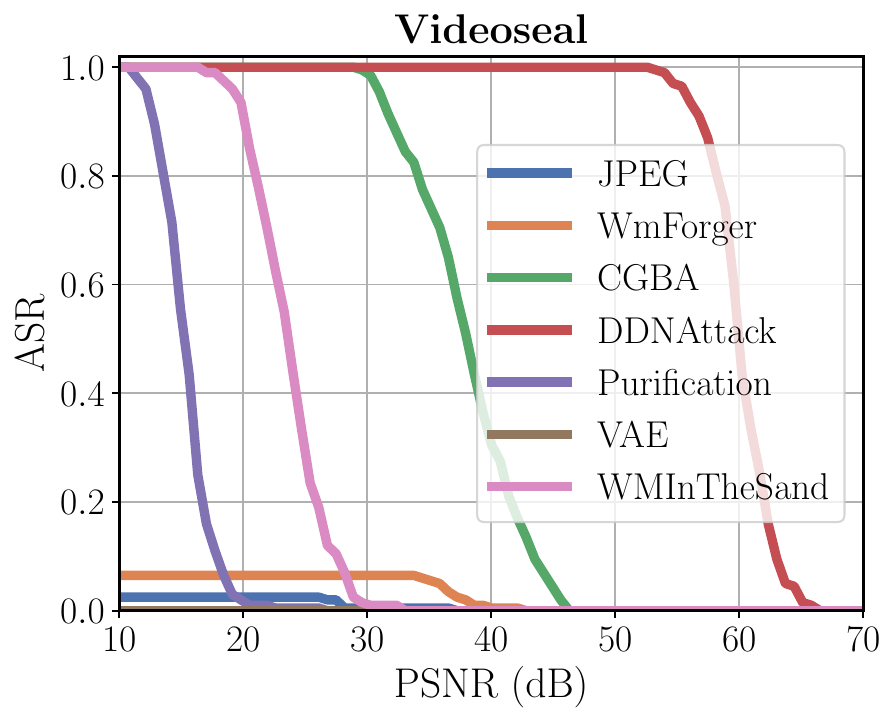}
    \includegraphics[width=0.45\linewidth]{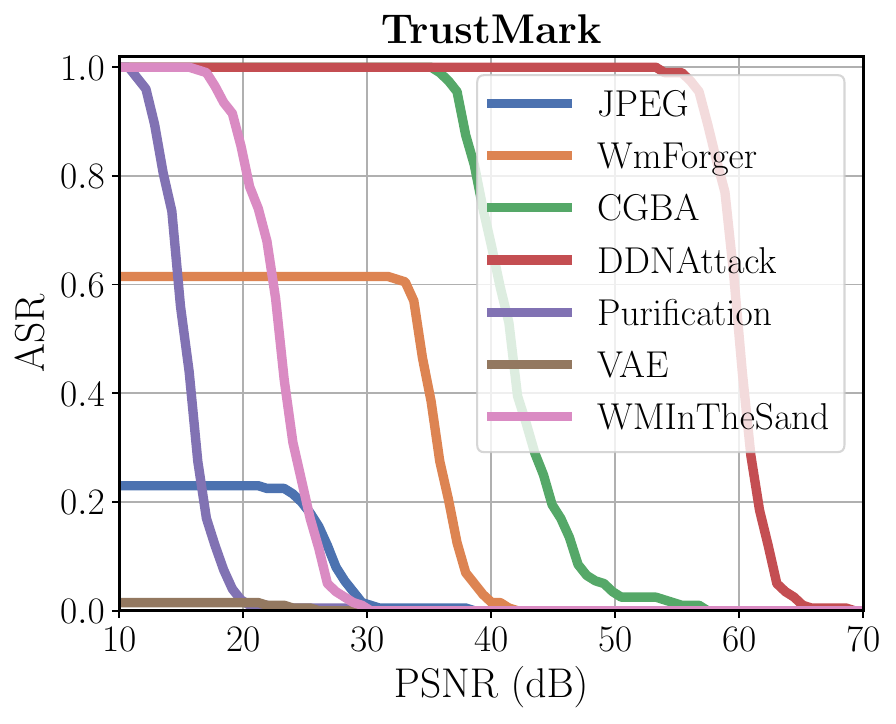}
    \includegraphics[width=0.45\linewidth]{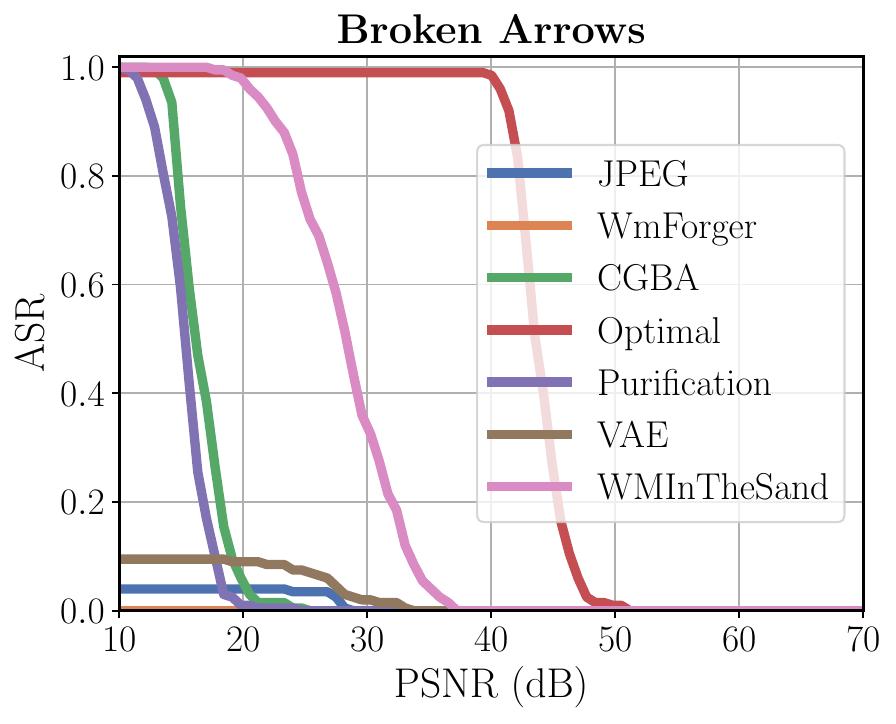}
    \caption{Comparison of attacks on Videoseal, TrustMark, and Broken-Arrows. Each curve represents the ASR as a function of the PSNR for a given attack. A smaller area is better for the watermark.}
    \label{fig:attacks}
\end{figure}

\paragraph{Oracle scenario} For lower distortion attacks ($PSNR \geq 35$ dB), Broken-Arrows is the most secure method. If we accept higher distortion, Videoseal becomes better. 
The figure~\ref{fig:attacks} shows the details of each attack depending on the watermarking methods. It shows that WmForger is the most efficient oracle attack against TrustMark, making it the least secure attack for $PSNR \geq 28$ dB.
This attack is less efficient against Videoseal and has no effect on Broken-Arrows watermarked images. 
Our WIS attack has a lot more impact on Broken-Arrows.
The VAE purification + downsampling operation results in systematic removal of the mid-frequency wavelet coefficients ($LH, HL, HH$), effectively filtering out the watermark signal embedded in those sub-bands. A possible defense is to concentrate the signal in the lowest frequency coefficients -- this would further improve robustness, but the resulting wavelet artifacts 

\paragraph{Blind Scenario} The results in Figure~\ref{fig:scenario} show the best performance of Regeneration, VAE Purification, and JPEG with a quality factor of 5, as it was the only classic value-metric augmentation that had an impact on the watermarking methods. 
In this scenario, the three methods are robust, with at least $PSNR \leq 30 $ dB required to erase the watermark. 
Notably, the Regeneration attack has the same efficiency across all methods.
A first difference between the methods lies in the VAE purification, which has a greater impact on Broken-Arrows than on modern methods for the same reasons as for WIS.
Secondly, TrustMark is not robust against strong JPEG compression, unlike Videoseal and Broken-Arrows. Even if Videoseal stands out as the most robust method in this scenario, the gain compared to Broken-Arrows is marginal. TrustMark is the overall worst choice of the three.

\section{Discussion and Conclusion}

Our findings reveal a critical trade-off between security and robustness when comparing modern and traditional watermarking. 
While modern DNN-based methods offer no significant gain in robustness under the tested conditions, they exhibit a significant decline in security. 
However, modern techniques are primarily optimized for high-capacity multi-bit payloads rather than the zero-bit identification scenario addressed here. 
Furthermore, this study does not account for geometric distortions, typically handled by modern methods.

\bibliographystyle{ACM-Reference-Format}
\bibliography{main,ML,Watermarking,Wm_Business,WM_Security,Adversarial} 

\appendix
\section{Our Watermark in The Sand}\label{app:wits}


\begin{algorithm}[t]
\caption{Watermark In The Sand (WIS) Attack}
\label{alg:wits_attack}
\begin{algorithmic}[1]
\Require Image $\mathbf{x}$, VAE $(\mathcal{E}, \mathcal{D})$, Oracle $\mathcal{O}$, Threshold $\beta$
\Ensure Adversarial Image $\mathbf{x}_A$

\State $\mathbf{x}_0 \gets \mathbf{x}$
\State $i \gets 0$

\Statex \textbf{// Phase 1: Iterative VAE Purification}
\While{$\mathcal{O}(\mathbf{x}_i) > \beta$}
    \State $\mathbf{x}_{down} \gets \text{Downsample}(\mathbf{x}_i, s)$ 
    \State $\mathbf{z} \gets \mathcal{E}(\mathbf{x}_{down})$ \Comment{Latent encoding}
    \State $\hat{\mathbf{x}}_{down} \gets \mathcal{D}(\mathbf{z})$ \Comment{Reconstruction}
    \State $\mathbf{x}_{i+1} \gets \text{Upsample}(\hat{\mathbf{x}}_{down}, \text{shape}(\mathbf{x}))$
    \State $i \gets i + 1$
\EndWhile

\Statex \textbf{// Phase 2: Patch-based refinement}
\State $\mathbf{x}_A, \mathbf{x}_B \gets \mathbf{x}_{i-1}, \mathbf{x}_i$
\While{$\mathcal{O}(\mathbf{x}_A) > \beta$}
    \State $m \gets \text{SelectPatchMask}()$
    \State $\mathbf{x}_A \gets (\mathbf{1} - m) \odot \mathbf{x}_A + m \odot \mathbf{x}_B$ \Comment{Patch replacement}
\EndWhile

\State \Return $\mathbf{x}_A$
\end{algorithmic}
\end{algorithm}

We developed a custom implementation of the WIS attack, adapting the original concept to improve visual fidelity and efficiency.
The original WIS method relies on an iterative patch-inpainting process governed by a GPT-as-a-Judge oracle.
This approach often introduces significant semantic distortions due to the inpainting and the stochastic nature of this type of oracle.
To mitigate these issues, our implementation introduces two key modifications.
First, we substitute the oracle with a distortion-based metric relative to the original image.
Second, we employ a VAE-purification process at the patch level.
The complete procedure is formalized in Algorithm~\ref{alg:wits_attack}.

\end{document}